# CLUSTER CORES, GRAVITATIONAL LENSING, AND COSMOLOGY

RICARDO A. FLORES
Department of Physics and Astronomy, University of Missouri, Saint Louis, MO 63121

and

JOEL R. PRIMACK
Santa Cruz Institute for Particle Physics, University of California, Santa Cruz, CA 95064

## Abstract

Many multiply–imaged quasars have been found over the years, but none so far with image separation in excess of 8″ . The absence of such large splittings has been used as a test of cosmological models: the standard Cold Dark Matter model has been excluded on the basis that it predicts far too many large–separation double images. These studies assume that the lensing structure has the mass profile of a singular isothermal sphere. However, such large splittings would be produced by very massive systems such as clusters of galaxies, for which other gravitational lensing data suggest less singular mass profiles. Here we analyze two cases of mass profiles for lenses: an isothermal sphere with a finite core radius (density $\rho \propto (r^2+r_{core}^2)^{-1}$), and a Hernquist profile ($\rho \propto r^{-1}(r+a)^{-3}$). We find that small core radii $r_{core} \sim 30h^{-1}$ kpc, as suggested by the cluster data, or large $a \gtrsim 300h^{-1}$ kpc, as needed for compatibility with gravitational distortion data, would reduce the number of large–angle splittings by an order of magnitude or more. Thus, it appears that these tests are sensitive both to the cosmological model (number density of lenses) and to the inner lens structure, which is unlikely to depend sensitively on the cosmology, making it difficult to test the cosmological models by large–separation quasar lensing until we reliably know the structure of the lenses themselves.

Gravitational lensing by foreground objects can produce multiple images of quasars and has been the subject of many analyses since the work of Turner, Ostriker, & Gott (1984). No multiply–imaged quasar is known with image separation in excess of 8″ , and only a few confirmed cases are known with splitting in excess of 3″ . Several studies (Narayan & White 1988, Cen et al. 1994, Wambsganss et al. 1994, Kochanek 1994) have concluded that the standard Cold Dark Matter (CDM) model predicts far too many large–angle splittings to be compatible with this fact, even if one takes into account that searches are biased against finding such systems (Kochanek 1994). In these studies the lenses are either modeled as singular isothermal sphere (SIS) halos, or assumed to be so at radii that cannot be resolved in N–body studies ($\lesssim 10h^{-1}$ kpc for a Hubble constant $H_0 = 100h$ kms$^{-1}$Mpc$^{-1}$). This assumption is amply justified in the study of small splittings because the lensing is due mostly to early type galaxies which indeed have very small core radii. A population of SIS lenses with the abundance of E/S0 galaxies adequately describes the small–splitting data (Kochanek 1993), and gives the line–of–sight probability distribution of image separations we show in Figure 1.

Large splittings ($\gtrsim 10''$ ) are due to much larger systems such as groups and clusters of galaxies for which the situation is different. In fact, there are several indications from gravitational lensing analyses that clusters have finite, albeit small, core radii $r_{core} \sim 20 - 30h^{-1}$ kpc. Tyson, Valdez, & Wenk (1990; hereafter TVW) studied the coherent alignment of background galaxies behind two rich clusters and found that $r_{core} > 20h^{-1}$ kpc is required, and profiles more singular than $r^{-1}$ at the center are also excluded by their data (Flores & Primack 1994; hereafter FP). The distortion of background images into radial "arcs" discovered in two rich clusters, MS 2137–23 (Fort et al. 1992) and A370 (Smail et al. 1995), shows that they cannot be described by singular potentials (Mellier, Fort, & Kneib 1993, Miralda–Escudé 1995). In the MS 2137–23 case, $r_{core} > 10h^{-1}$ kpc*, but the data favor a larger value $r_{core} \sim 30h^{-1}$ kpc. In the case of A370 the data favor a core radius $\sim 25h^{-1}$ kpc, but we do not know if this corresponds to an isothermal density profile. In general, good fit lens models of arc+arclet fields require $r_{core} \sim 20 - 30h^{-1}$ kpc, and we are not aware of any good fit with a singular lens. Furthermore, while the abundance of giant arcs in clusters has been argued to require singular cluster potentials (see Wu & Hammer 1993 and references therein), Bartelmann et al. (1994) have shown that if clusters are asymmetric and have a significant amount of substructure, as the data indicate (see Bird 1994, Struble & Ftaclas 1994, and references therein), even clusters with core radii as large as $r_{core} \sim 50h^{-1}$ kpc generate such arcs as efficiently as clusters modeled as smooth SIS lenses. Finally, Bergmann & Petrosian (1993) have shown that the observed small proportion of long arcs to arclets is inconsistent with SIS lenses, but it is sensitively dependent on core radius and consistent with $r_{core} \sim 50h^{-1}$ kpc.

Core radii $r_{core} \gtrsim 20 - 30h^{-1}$ kpc have a drastic effect on the probability of multiple imaging because the cross section for multiple imaging vanishes for a spherically symmetric lens of core radius $r_{core} \geq 33(v/1000\mathrm{km\,s^{-1}})^2 h^{-1}$ kpc, where $v$ is the one–dimensional velocity dispersion, if the source is at redshift $z_S \leq 3$ (Hinshaw & Krauss 1987). Kochanek (1994) has briefly considered the effect of a non–zero core radius on the frequency of large angle splittings as a possible systematic uncertainty, but he points out that the known large separation systems do not show the central image that would be expected if the lens were a cored

* We quote the lower bound of Miralda–Escudé (1995) because the constraint of Mellier, Fort, & Kneib (1993) was derived for a different potential.

isothermal sphere. However, the well studied large separation systems do not seem to be generated by a single cluster–size lens. In the case of Q0957+561 (separation $6.1''$ ) the lensing is clearly produced by a giant elliptical in conjunction with the cluster (see Dahle, Maddox, & Lilje 1994, and references therein). In the case of Q2016+112 (separation $3.6''$ ) recent observations also indicate that multiplane lensing is more likely (Garrett et al. 1994). Finally, for Q2345+007 (separation $7.1''$ ) there is a galaxy very close to one of the quasar images and two clumps of galaxies farther away (Mellier et al. 1994, Fisher et al. 1994), making it also a rather complex system of lenses. Thus, there is no indication that we have seen multiple imaging by a single cluster–size lens as yet.

In this *Letter* we consider the effect of two kinds of non–SIS lenses in the analyses of the expected frequency of large–splitting quasar lensing in cosmological models: a cored isothermal sphere (CIS) $\rho \propto (r^2+r_{core}^2)^{-1}$, and a Hernquist profile, $\rho \propto r^{-1}(r+a)^{-3}$ (Hernquist 1990; see FP for discussion of our motivations for considering this, and Navarro, Frenk, & White (1995) regarding latest results from simulations). We reconsider the recent studies of Cen et al. and Wambsganss et al. and quantify the expected change in the computed frequencies of large–angle splittings if the lenses were assumed to have these density profiles. We find that the computed frequencies decrease by an order of magnitude or so once the changes in lensing cross section *and in magnification bias* are taken into account. Including the latter is crucial to avoid misleading results.

We begin by calculating the line–of–sight angular probability distribution for SIS lenses as a function of the abundance of the lenses, which we then compare to the results of Cen et al. (1994) in order to fix the abundance and calculate the change in the probability distribution if one assumes non–SIS lenses. We shall assume spherically symmetric lenses because the lensing cross section for multiple imaging of a *point* source is not likely to be sensitive to substructure and asphericity since it does not depend differentially on the light bending angle, $\alpha$. As we explain below, the image splitting $\theta$ is determined by the lens equation

$$\alpha(x) = \frac{D_{OS}\ x}{D_{OL} D_{LS}} - \frac{D_{OS}}{D_{LS}}\phi\ , \tag{1}$$

where $D_{OL}, D_{OS}$, and $D_{LS}$ are the observer–lens, observer–source, and lens–source angular–diameter distances respectively, $x$ is the impact parameter of the light ray, $\phi$ is the unperturbed angular position of the source, and $\alpha(x) = 4GM(|x|)/c^2 x$ for a projected mass $M(|x|)$ inside radius $|x|$. For an isothermal sphere, $\alpha(x) = N\left(\sqrt{x^2 + r_{core}^2} - r_{core}\right)/x$ (Hinshaw & Krauss 1987), where $N = 4\pi v^2/c^2$ if one normalizes the non–singular profile to asymptotically enclose the same mass as the SIS profile. We discuss our normalization below. For a Hernquist profile we find

$$\alpha(x) = N' \frac{x}{a}\ \frac{\log\left(\frac{\sqrt{a+|x|}+\sqrt{a-|x|}}{\sqrt{a+|x|}-\sqrt{a-|x|}}\right) - \sqrt{1-(x/a)^2}}{\left(1-(x/a)^2\right)^{3/2}}\ ; \quad |x| \leq a \tag{2a}$$

$$\alpha(x) = N' \frac{x}{a}\ \frac{2\tan^{-1}\left(\frac{\sqrt{|x|+a}}{\sqrt{|x|-a}}\right) + \sqrt{(x/a)^2-1} - \pi}{\left((x/a)^2-1\right)^{3/2}}\ ; \quad |x| \geq a \tag{2b}$$

There are three solutions ($x_i$) to Equation (1) in general, provided that the source is within angular position $\phi = \phi_c$ at which the right–hand side is tangent to $\alpha(x)$. Thus, the lensing cross section is $\pi(D_{OL}\phi_c)^2$. The image separation between the two most–split images is $\theta = (D_{LS}/D_{OS})(\alpha(x_1) - \alpha(x_3))$, and is fairly insensitive to $\phi$ (we typically find $|\theta(\phi)/\theta(\phi = 0) - 1| \lesssim 0.1$ for $\theta$ in the range $10'' - 100''$ and $r_{core} \lesssim 30h^{-1}$ kpc); therefore we shall approximate $\theta = 2(D_{LS}/D_{OS})\alpha(x_1(\phi = 0))$.

The differential probability for a beam to encounter a lens at redshift $z_L$ that will cause a splitting $\theta$ when traversing the path $dz_L$ is (see e.g. Fukugita et al. 1992)

$$\frac{dP}{d\theta} = \frac{dn(\theta, z_L)}{d\theta} \sigma(\theta, z_L) \frac{cdt}{dz_L} dz_L \ , \tag{3}$$

where $dn(\theta, z_L)/d\theta$ is the comoving number density of lenses capable of producing a splitting within $d\theta$ of $\theta$, $\sigma(\theta, z_L)$ is the lensing cross section, and $cdt/dz_L = (c/H_0)/(1+z_L)^2\sqrt{1+\Omega z_L}$ with $\Omega$ the mean density in units of the critical density. We shall assume an unevolving population of lenses, so that $dn(\theta, z_L)/d\theta = (1+z_L)^3 dn(\theta, 0)/d\theta$; this is a reasonably good approximation to the numerical results of Cen et al. (1994) (see their Fig. 6 and discussion therein). We shall assume a Schechter form for the comoving density of lenses of a given virial mass $M$, $n(M) = n_* M_*^{-1}(M/M_*)^\alpha \exp(-M/M_*)$, and a relation $M/M_* = (v/v_*)^\gamma$ between $M$ and the dispersion $v$. We then fix the parameters to values that allow us to fit the distribution $dP/d\theta$ obtained from numerical simulations by Cen et al. (1994).

We show $dP/d\theta$, Equation (3), integrated in angular bins in Figure 1. The shape of the angular distribution is independent of redshift under our assumption of no evolution. We choose to fit the Cen et al. (1994) data for a source at redshift $z_S = 2$ since this is roughly the redshift at which the observed redshift distribution of quasars peaks. Assuming SIS lenses, a reasonable fit (solid squares) to the shape of the Cen et al. (1994) angular distribution is obtained for $v_* = 870$ kms$^{-1}$, $\gamma = 3$, and $\alpha = -1.5$. The fit to the amplitude gives $n_* \approx 2.5 \times 10^{-4} h^3$ Mpc$^{-3}$, about 10 times the mean density of Abell clusters. This is a well known problem of CDM (see e.g. White, Efstathiou, & Frenk 1993). Also, the high–dispersion tail of the distribution is far sharper than that of observed groups and clusters, another feature of CDM that appears to be a problem (see Zabludoff & Geller 1994, and references therein). The fall–off of the integrated probability $P(>\theta)$ at large $\theta$ is not sharp enough to match the fall–off calculated beyond $\theta \gtrsim 40''$ by Wambsganss et al. (1994), but this is not a crucial shortcoming of the fit for our purposes. If we restrict the images to brightness difference $\Delta m \leq 1.5$ mag, as Wambsganss et al. (1994) do, we get $P(\theta \geq 10'') = 0.0004(0.0011, 0.0018)$ for a source at $z_S = 1(2, 3)$, compared to $0.0007(0.0014, 0.002)$ in Wambsganss et al. (1994). Thus, the fit is reasonably good for our purpose of quantifying the changes in the lensing probabilities of Cen et al. (1994) and Wambsganss et al. (1994) for non–SIS lenses.

We show $dP/d\theta$ for the non–SIS lenses considered here, integrated over the same angular bins as above, in Figure 1. In a previous study (FP) we found that for a Hernquist profile, only large $a$ is compatible with the distortion of background galaxies by a cluster measured by TVW. Therefore, we consider $a = 10\, r_{core}$. We also consider a smaller value, $a = r_{core}$, for comparison. We fix the normalization factors $N$ and $N'$ above to give equal projected mass within a radius $r_N = 100h^{-1}$ kpc, so that the profiles adequately approximate the projected mass distribution of the lenses for radii in the range $(50 - 100)h^{-1}$ kpc, and deviate significantly from a SIS profile only in the inner part where the numerical simulations cannot resolve the mass distribution. As can be seen in Figure 1, the probabilities are reduced by about two orders of magnitude for $r_{core} = 30h^{-1}$ kpc and for the Hernquist profile with $a = 10\, r_{core} = 300h^{-1}$ kpc, but by a smaller factor for $a = 30h^{-1}$ kpc. This does not translate into an equal reduction in the expected number of lensed quasars, however, because observations have a rather limited dynamic range and lensed quasars are more likely to be found due their increased brightness (Kochanek 1994, Fukugita & Turner 1991, and references therein). The

line–of–sight probabilities of Wambsganss et al. (1994) are reduced by nearly this factor nonetheless, because restricting the dynamic range to $\Delta m < 1.5$ mag does not significantly reduce the cross section for $\theta \gtrsim 10''$ . In Table 1, we give the line–of–sight probability, $P(\Delta m < 1.5\,\mathrm{mag}, \theta > 10'')$, relative to the SIS value.

Cen et al. (1994) do not restrict the dynamic range to a range accessible to observations, but do incorporate the magnification bias due to increased brightness in order to predict the number of lensed systems. They find that if one requires that even the fainter image of a lensed quasar be brightened enough to be observable, CDM predicts that about 7 or 8 lensed quasars with image separation in excess of $8''$ should be present in a sample of the size of the Hewitt & Burbidge (1989; hereafter HB) catalog (and many more if only the brighter image is required to be observed). Since none are known, this rules out the CDM model assuming SIS lenses. In Table 2, we give the line–of–sight probability corrected for the faint bias, relative to the SIS value corrected for the faint bias, both calculated for the quasars in the HB catalog so that this reduction factor can be directly applied to the CDM predictions of Cen et al. (1994). Notice that the bias correction can be very large for non–SIS models, hence the much increased ratios relative to Figure 1. The value of the core radius is assumed to be the same for all lenses here (see the discussion below), and the count of quasars of apparent magnitude $m$ fit to the usual broken power–law, $dN/dm \propto 10^{\alpha(m-m_0)}(10^{\beta(m-m_0)})$ for $m < m_0 (m > m_0)$, with $(\alpha, \beta, m_0) = (0.86, 0.28, 19.15\,\mathrm{mag})$ (see Fukugita & Turner 1991). As can be seen in Table 2, the predicted number of lensed quasars with $\theta \geq 8''$ is drastically reduced, enough so to make the number compatible with observations for $r_{core} \gtrsim 20h^{-1}$ kpc, except for the Hernquist profile with $a = r_{core}$.

These numbers do not fully reflect the observational selection effects, however,

because they do not include the limited dynamic range of observations. The magnification bias is most significant for quasars brighter than the break magnitude $m_0$, and Kochanek (1994) has recently shown that bright quasars are likely to be detected very close to the magnitude limit of the search. For a quasar of magnitude 18.5, roughly the magnitude at which the distribution of quasars in the HB catalog peaks, the median dynamic range is only 0.37 mag. Thus, it is unlikely that images with amplification ratios of more than about 1.5 would be detected. Including this restriction would decrease the number of lensed quasars predicted. In addition, Kochanek (1994) notes that more recent studies indicate a steeper luminosity function at the bright end, $\alpha = 1.12$, which would increase the magnification bias. In Table 3, we give these numbers assuming a uniform dynamic range corresponding to the median at magnitude 18.5 and for $\alpha = 1.12, \beta = 0.18$. One can see that the two effects more or less cancel each other out, and the numbers remain small.

These results show that lenses with core radii $r_{core} \gtrsim 30h^{-1}$ kpc would reduce the expected number of large–separation lenses to levels compatible with observations, and perhaps even values as small as $r_{core} = 20h^{-1}$ kpc might do so if one considers that the current sample is probably not more than 20% complete (Kochanek 1994). We have assumed the lenses to have a common core radius, although we do not have any information on the core radii of groups of galaxies. However, most of the contribution to the line–of–sight probability of large splitting, $\theta \geq 10''$ , of sources at redshift $z_S \leq 2$ comes from lenses with large velocity dispersion, $v \gtrsim 900(1000)$ km s$^{-1}$ for $r_{core} \geq 20(30)h^{-1}$ kpc, therefore this assumption is self–consistent. We have nonetheless explored the hypothesis of a variable core radius, $r_{core} = r_0(v/1000\mathrm{km\,s^{-1}})^2$ with $r_0 = 20 - 30h^{-1}$ kpc. This scaling is consistent with an analysis of the dynamics of the globular clusters in M87, which suggests $r_{core}^{M87} \sim 6$ kpc (Merritt & Tremblay 1993), and it is enough to show how variable core radii would affect our results. We find that the ratios of Table 3 typically *decrease* if we assume this scaling; e.g. the ratio of probability of splitting in excess of $8''$ for CIS lenses is reduced to $0.039(0.00063, 0.32)$ for $r_0 = 25(30, 20)$ kpc if we restrict $r_{core} \le 100h^{-1}$ kpc, and even more if we do not. Our results depend on the normalization radius $r_N$, because at a fixed image separation a rising (falling) $\alpha(x)$ requires a more (less) massive lens within $r_N$ than a SIS lens , which yields a smaller (larger) probability. Since we use an unevolving population of lenses, and quasars are mostly at redshifts $z_S \lesssim 3$, it is the mass that has virialized by redshift 3 that determines the number density of lenses that can produce a given separation. A cluster–size perturbation is virialized inside a radius $\approx 200(v/1000\mathrm{km\,s}^{-1})h^{-1}$ kpc by redshift 3 (which is why the Cen et al. (1994) results are fairly well approximated with an unevolving population of lenses, because for $\theta \le 100''$ the impact parameter $x_1(\phi = 0) \lesssim 200h^{-1}$ kpc for sources at redshift $z_S \le 3$), but we have used $r_N = 100h^{-1}$ kpc to get a conservative (*i.e.* high) estimate of $dP/d\theta$. The ratios of Table 3 typically *decrease* if we use $r_N = 200h^{-1}$ kpc; e.g. the ratio of probability of splitting in excess of $8''$ for CIS lenses is reduced to $0.087(0.26)$ for $r_{core} = 30(20)h^{-1}$ kpc.

We have assumed spherical lenses here, but this should not be a serious limitation. Ellipticity will not change the average cross section, and the magnification bias is in fact not very different from the circular case (Kochanek 1994). Substructure might affect the magnification bias, but less so our reduction factors because the effect would partially cancel out in the ratio. Therefore the crucial question is whether substructure is adequately taken into account in the numerical work.

Kochanek (1994) cannot include substructure in his calculations. The ray–tracing technique used by Wambganss et al (1994) would include effects due to subtruc- ture, but overmerging in numerical simulations may make the lenses unrealistically smooth (see e.g. Moore, Katz, & Lake 1995).

The absence of lensed quasars with large image separation has been found incompatible with the predictions of a CDM $\Omega = 1$ cosmology assuming that lenses are singular isothermal spheres (SIS). We have studied the probability of large–separation lensing of quasars in a CDM $\Omega = 1$ universe assuming various non-SIS mass profiles for lenses. We have found that if the cluster–size lenses that can generate large splittings in quasar images are either (a) isothermal spheres with small, but finite, core radius $r_{core} \sim 20 - 30h^{-1}$ kpc, as indicated by other gravitational lensing data, or (b) spheres with a Hernquist density profile of large $a \gtrsim 300h^{-1}$ kpc, as required for a Hernquist profile by measurements of the lensing distortion of distant galaxies, and for which the density profile changes outward from $\rho \propto r^{-1}$ to $\rho \propto r^{-2}$ inside a radius $\sim 300h^{-1}$ kpc, then the expected number of such splittings is small enough to be compatible with their absence in the present data. These results have several implications: (1) The large–separation quasar lensing test is sensitive both to the cosmological model (mostly the number density of lenses) and to the inner lens structure, which is unlikely to depend sensitively on the cosmology, making it difficult to probe the models by this test until we reliably know the structure of the lenses. (2) To the extent that the problem for CDM in this context is only that it predicts about ten times the observed density of clusters at the present, rather than SIS profiles, and evolution of cluster cores for $z_L \lesssim 0.5$ is not important, these results indicate that it will be difficult to find lensed quasars with large splittings. (3) If clusters and groups were indeed nearly singular and unevolving for $z_L \lesssim 0.5$, we could expect a few large–separation images to turn up if all the quasars in a quasar catalog as large as the HB catalog were searched for an accompanying image. But much larger catalogs will soon become available, for example as a result of the Sloan Digital Sky Survey. Thus it would be worth calculating the number of wide-separation lensed quasars that would be predicted in various cosmological models which, unlike COBE-normalized CDM, fit the observed number densities of clusters and groups, for various assumptions about their inner density profiles.

We thank Renyue Cen for explaining to us the bias factor for the HB catalog used in Cen et al. (1994), and Chris Kochanek for helpful comments. This work has been partially supported by fellowships from Fundación Andes and UM–St. Louis (RF), and NSF grants.

## TABLE CAPTIONS

**Table 1** – Line–of–sight probability of image splitting in excess of $10''$ and magnitude difference between images of less than 1.5 mag for a source at redshift $z_S = 1, 2$ and 3. The values are given relative to the probability with the same restrictions but for SIS lenses. Three kinds of lenses are considered: CIS lenses with $r_{core} = 30$ and $20h^{-1}$ kpc, and Hernquist lenses with $a = r_{core}$ and $a = 10r_{core}$.

**Table 2** – Line–of–sight probability, corrected for magnification bias of the fainter image, for the most–split images to have a separation $\theta$ in the given bins. Same lenses as in Table 1 are considered. The magnification bias is calculated for the HB catalog and with a quasar magnitude count with parameters $(\alpha, \beta, m_0) = (0.86, 0.28, 19.15\text{mag})$. The values are given relative to the probability assuming SIS lenses. (We determine the range of $\phi$ over which the magnification for each cluster mass profile exceeds a given value $\mu$, and determine the magnification probability distribution $P(> \mu)$ from the total area calculated numerically.)

**Table 3** – Line–of–sight probability, corrected for magnification bias of the fainter image, for the most–split images to have a separation $\theta$ in the given bins and magnitude difference of less than 0.37 mag, the median value at magnitude 18.5 mag for a quasar magnitude count with parameters $(\alpha, \beta, m_0) = (1.12, 0.18, 19.15\text{mag})$. Same lenses as in Table 1 are considered. The magnification bias is calculated for the HB catalog, and the probability values are given relative to the probability assuming SIS lenses.

**TABLE 1**

| **P($\Delta$m < 1.5 mag, $\theta > 10''$ ) Relative to SIS** | | | | |
|---|---|---|---|---|
| redshift | Profile | | | |
| $z_S$ | SIS | $r_{core} = 30(20)h^{-1}$kpc | $a = r_{core}$ | $a = 10r_{core}$ |
| 1 | 1 | 0.0053(0.017) | 1.2(2.6) | 0.013(0.038) |
| 2 | 1 | 0.016(0.042) | 1.5(2.6) | 0.032(0.072) |
| 3 | 1 | 0.023(0.054) | 1.5(2.6) | 0.041(0.089) |

**TABLE 2**

| **Faint Bias Probability Relative to SIS** | | | | |
|---|---|---|---|---|
| $\theta$ bin | Profile | | | |
| arcsec | SIS | $r_{core} = 30(20)h^{-1}$kpc | $a = r_{core}$ | $a = 10r_{core}$ |
| $8 - 16$ | 1 | 0.0088(0.040) | 0.44(0.84) | 0.044(0.12) |
| $16 - 32$ | 1 | 0.069(0.19) | 0.83(1.2) | 0.15(0.32) |
| $32 - 64$ | 1 | 0.29(0.48) | 0.89(1.2) | 0.39(0.58) |
| $> 8$ | 1 | 0.10(0.20) | 0.69(1.1) | 0.17(0.30) |

**TABLE 3**

| **Faint Bias Probability, for $\Delta$m < 1.5 mag, Relative to SIS** | | | | |
|---|---|---|---|---|
| $\theta$ bin | Profile | | | |
| arcsec | SIS | $r_{core} = 30(20)h^{-1}$kpc | $a = r_{core}$ | $a = 10r_{core}$ |
| $8 - 16$ | 1 | 0.0062(0.048) | 0.62(0.94) | 0.094(0.27) |
| $16 - 32$ | 1 | 0.11(0.40) | 0.87(1.2) | 0.36(0.73) |
| $32 - 64$ | 1 | 0.64(1.1) | 0.85(1.0) | 0.89(1.3) |
| $> 8$ | 1 | 0.20(0.42) | 0.75(1.1) | 0.39(0.68) |

## FIGURE CAPTIONS

**Figure 1** – Line–of–sight probability of image splitting by an angle $\theta$ for various logarithmic $\theta$–bins. The differential probability, $dP/d\theta$, integrated over the range of the angular bins, is plotted as a function of the binned separation angle of the most–split images, $\theta$. The dotted histogram is the numerical data of Cen et al. (1994; see their Figure 3$a$) assuming SIS lenses, and for a source at redshift $z_S = 2$ in a CDM, $\Omega = 1$ universe. The solid squares are our Schechter–type fit to these data. The solid (dashed) histogram is the probability recalculated with CIS (Hernquist) lenses. For the CIS case with $r_{core} = 30h^{-1}$ kpc and the Hernquist profile with $a = 10r_{core} = 300h^{-1}$ kpc, the lensing probability is reduced very substantially. Finally, the open squares represent the angular distribution of a model that can account for the observed small–separation lenses (see Kochanek 1993 for details).

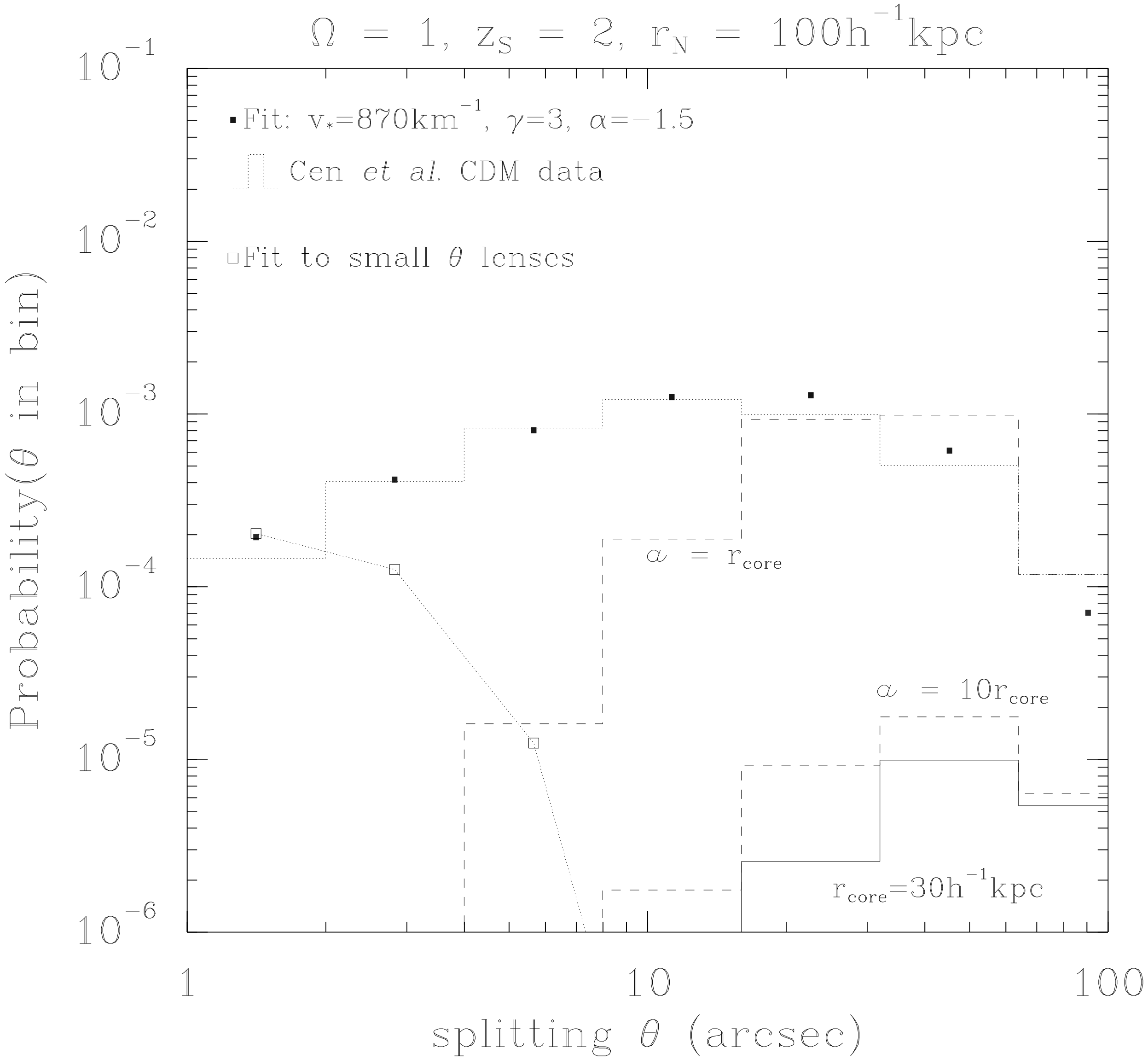

Ω = 1, z_S = 2, r_N = 100h^-1kpc
Fit: v*=870km^-1, γ=3, α=−1.5
Cen et al. CDM data
Fit to small θ lenses
α = r_core
α = 10r_core
r_core=30h^-1kpc
Probability(θ in bin)
splitting θ (arcsec)
10^-1
10^-2
10^-3
10^-4
10^-5
10^-6
1
10
100